\documentclass[aps,pre,reprint,amsmath,amssymb,superscriptaddress,showpacs,floatfix]{revtex4-1}
\usepackage{graphicx}
\usepackage{dcolumn}
\usepackage{bm}
\usepackage[colorlinks, allcolors=blue]{hyperref}
\usepackage[usenames, dvipsnames]{color}

\begin{document}
\title{Bond-bond correlations, gap relations and thermodynamics of spin-$1/2$ chains with spin-Peierls transitions and bond-order-wave phases}

\author{Sudip Kumar Saha}
\affiliation{S. N. Bose National Centre for Basic Sciences, Block - JD, Sector - III, Salt Lake, Kolkata - 700106, India}

\author{Manoranjan Kumar}
\email{manoranjan.kumar@bose.res.in}
\affiliation{S. N. Bose National Centre for Basic Sciences, Block - JD, Sector - III, Salt Lake, Kolkata - 700106, India}

\author{Zolt\'an G. Soos}
\email{soos@princeton.edu}
\affiliation{Department of Chemistry, Princeton University, Princeton, New Jersey 08544, USA}

\date{\today}

\begin{abstract}

The spin-$1/2$ chain with antiferromagnetic exchange $J_1$ and $J_2 = \alpha J_1$ between first and second neighbors, respectively, 
has both gapless and gapped ($\Delta(\alpha) > 0$) quantum phases at frustration $0 \le \alpha \le 3/4$. The ground state instability of regular 
($\delta = 0$) chains to dimerization ($\delta > 0$) drives a spin-Peierls transition at $T_{SP}(\alpha)$ that varies with $\alpha$ in these strongly 
correlated systems. The thermodynamic limit of correlated states is obtained by exact treatment of short chains followed by density matrix 
renormalization calculations of progressively longer chains. The doubly degenerate ground states of the gapped regular phase are bond order waves (BOWs) 
with long-range bond-bond correlations and electronic dimerization $\delta_e(\alpha)$. The $T$ dependence of $\delta_e(T,\alpha)$ 
is found using four-spin correlation functions and contrasted to structural dimerization $\delta(T,\alpha)$ at $T \le T_{SP}(\alpha)$. 
The relation between $T_{SP}(\alpha)$ and the $T = 0$ gap $\Delta(\delta(0),\alpha)$ varies with frustration in both gapless and gapped 
phases. The magnetic susceptibility $\chi(T,\alpha)$ at $T > T_{SP}$ can be used to identify physical realizations of spin-Peierls systems. 
The $\alpha = 1/2$ chain illustrates the characteristic BOW features of a regular chain with a large singlet-triplet gap and electronic dimerization.  

\end{abstract}


\pacs{}

\maketitle


\section{\label{sec1}Introduction}

The Peierls instability of polyacetylene is associated in chemistry with bond length alternation of polyenes~\cite{longuet59} and in physics 
with topological solitons and mid-gap excitations of the Su-Schrieffer-Heeger model~\cite{su1979, su1980}. In terms of a half-filled H\"uckel 
or tight-binding band, the ground state of the regular ($\delta= 0$) polymer with equal C-C bond lengths is unconditionally unstable against a 
harmonic ($\delta^2$) potential. Peierls instabilities and transitions have been studied in many materials with quasi-1D chains~\cite{jerome2004,pouget2017,special_neural}, 
both inorganic and organic, both conductors and insulators in the high-$T$ phase with $\delta = 0$ and equal exchange $J_1$ or electron 
transfer $t$ between neighbors along the chain. The simplest systems are spin-$1/2$ chains with two degrees of freedom per site and spin-Peierls transitions 
at $T_{SP}$ to a structure with lower symmetry. The chains discussed in this paper have alternating exchanges $J_1(1 \pm \delta(T)$) and dimerization $\delta(T)$ 
that decreases with $T$ and vanishes at $T_{SP}$.

Exotic quantum phases of 1D models with frustrated or competing interactions are a related topic of current interest~\cite{hikihara2008,sudan2009}. 
The many body problem is typically defined on a regular ($\delta= 0$) chain and addressed by multiple theoretical and numerical methods. 
As noted by Allen and S\'en\'echal~\cite{allen1997}, spin-$1/2$ chains are either gapless with a nondegenerate ground state or gapped with a doubly degenerate 
ground state. The $J_1-J_2$ model discussed below has isotropic antiferromagnetic exchanges $J_1$ and $J_2 = \alpha J_1$ between first and second 
neighbors. The quantum critical point~\cite{nomura1992} $\alpha_c = 0.2411$ separates gapless and gapped phases with increasing frustration $\alpha$. The ground states 
refer to rigid chains that exclude Peierls transitions or, indeed, physical realizations. Double degeneracy indicates a bond-order-wave (BOW) phase 
with spontaneously broken inversion symmetry at sites in the ground state. As Nakamura~\cite{nakamura2000} predicted, half-filled extended Hubbard~\cite{sengupta2002} and 
related~\cite{mkumar2009} models 
support BOW phases in narrow ranges of parameters. The dimer phase of the $J_1-J_2$ model has a BOW ground state.

We have recently successfully modeled~\cite{sudipsp2020} the best characterized SP systems, the organic crystal~\cite{jacob1976} TTF-CuS$_4$C$_4$(CF$_3$)$_4$ with $\alpha= 0$ 
and the inorganic crystal~\cite{haseprb1993,uchinokura2002} CuGeO$_3$ with $\alpha= 0.35$. We study in this 
paper the $J_1-J_2$ model, Eq.~\ref{eq:j1j2_dim} below, with linear coupling to a 
harmonic lattice and variable frustration $0 \le \alpha \le 3/4$ that includes both gapped and gapless phases. We distinguish between phase transitions 
with structural dimerization $\delta(T,\alpha)$ at $T < T_{SP}(\alpha)$ and quantum transitions with electronic dimerization $\delta_e(0,\alpha)$ 
for $\alpha > \alpha_c$ in a rigid lattice, and we obtain the $T$ dependence of $\delta_e(T,\alpha)$ using bond-bond correlation functions.

\begin{figure}[t]
	 \begin{center} \includegraphics[width=\columnwidth]{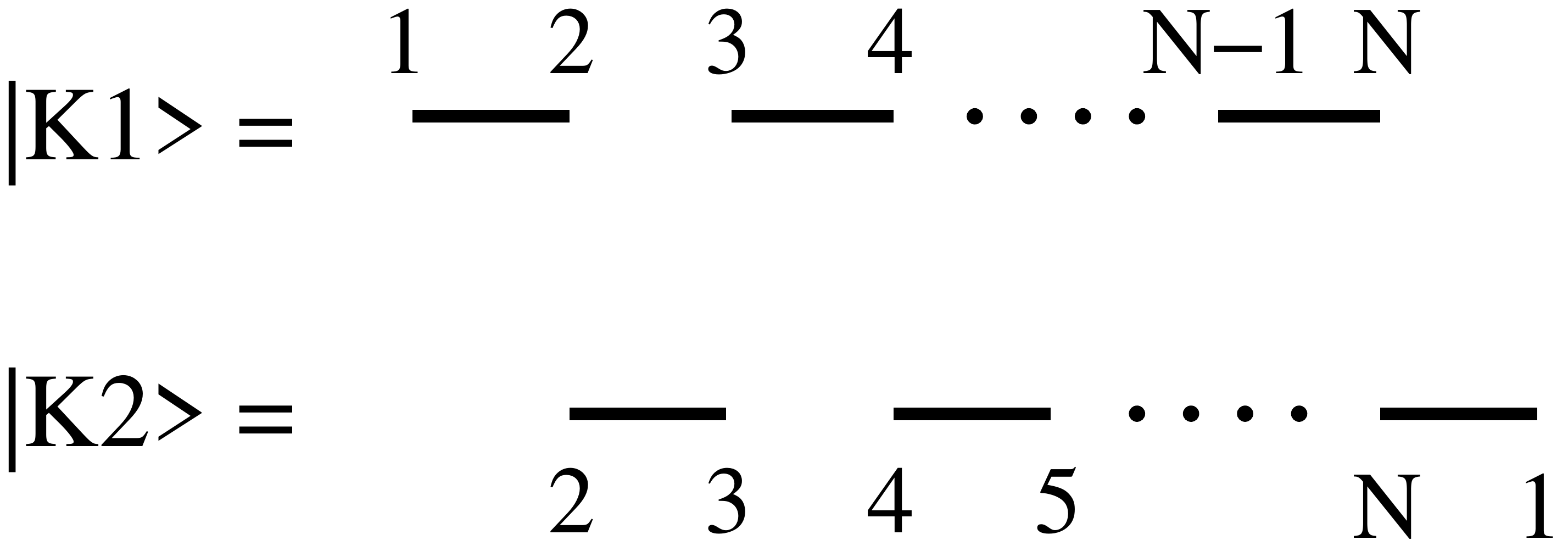}
 \caption{
	 Kekul\'e valence bond diagrams. Lines indicate singlet-paired spins in Eq.~\ref{eq:kekule}.} 
 \label{fig1}
 \end{center}
 \end{figure}

The following special case contrasts electronic and structural dimerization. At the Majumdar-Ghosh (MG) point~\cite{ckm69b}, $\alpha = 1/2$, 
the exact ground states for an even number of spins $N$ are the Kekul\'e  valence bond (VB) diagrams $\vert K1 \rangle$ and $ \vert K2 \rangle$ sketched 
in Fig.~\ref{fig1}. Each diagram is a product of $N/2$ singlet-paired spins shown as lines
\begin{eqnarray}
	\left(r,r+1\right) &= \left( \alpha_r \beta_{r+1}-\beta_r \alpha_{r+1} \right) / \sqrt{2}.
\label{eq:kekule}
\end{eqnarray}
$\vert K1 \rangle$ has odd $r$ and paired spins $(1,2)(3,4)...(N-1,N)$ while $\vert K2 \rangle$ has even $r$ and paired spins $(2,3)(4,5)...(1,N)$. 
Either diagram has perfectly ordered bonds. The linear combinations $\vert K1 \rangle \pm \vert K2 \rangle$ are even and odd, respectively, 
under inversion at sites and also have long-range bond-bond order. 

 The linear Heisenberg antiferromagnet (HAF) has $\alpha= 0$ and $J_1(1 \pm \delta)$ at dimerization $\delta$. 
 The limits $\delta= \pm 1$ lead to $N/2$ two-level systems with singlet ground states $\vert K1 \rangle$ and $\vert K2 \rangle$ 
 for odd and even $r$ in Eq.~\ref{eq:kekule}. Since the \textit{nondegenerate} ground state of the $\delta= \pm 1$, $\alpha = 0$ chains have 
 perfectly ordered bonds, just as do the \textit{degenerate} ground states of the $\delta = 0$, $\alpha = 1/2$ chain, we conclude that 
 $\delta_e = \pm 1$ at the MG point. Any observable associated with the electronic ground state must be identical. The energy spectrum 
 is otherwise quite different. Bond-bond correlation functions in Section~\ref{sec3} generalize $\delta_e(\alpha)$ to frustration 
 $\alpha$ in the ground state and then to $\delta_e(T,\alpha)$.

In this paper, we compute the thermodynamics of finite chains with periodic boundary conditions using exact diagonalization (ED) for short 
chains and the density matrix renormalization group (DMRG) for increasingly long chains~\cite{sudip19}. DMRG~\cite{white-prl92,white-prb93} is a state-of-the-art 
numerical technique to deal with strongly correlated 1D electronic systems. We use the modified DMRG algorithm for systems with periodic 
boundary conditions~\cite{ddpbc2016}. The details of the ED/DMRG method are discussed in Ref. \cite{sudip19}. The basic idea is that finite $T$ 
limits the range of correlations in general, while a finite gap limits the range at $T = 0$. Dimerization opens a gap in gapless chains 
and increases the gap $\Delta(\alpha)$ of the $J_1-J_2$ model with $\alpha > \alpha_c$. It follows that the thermodynamic limit is reached 
at a system size that depends on $T_{SP}$ in deformable chains or on $\Delta(\alpha)$ in rigid regular chains. 

The paper is organized as follows. The $J_1-J_2$ model on deformable chains is defined in Section~\ref{sec2} along with the 
equilibrium condition for dimerization $\delta(T,\alpha)$ and criterion for the thermodynamic limit. Bond-bond correlation 
functions are used in Section~\ref{sec3} to discuss electronic dimerization $\delta_e(T,\alpha)$ in rigid regular chains with 
BOW ground states. Gap relations of $J_1-J_2$ models with SP transitions are reported for gapped and gapless phases in Section~\ref{sec4} 
and contrasted to free fermions. The magnetic susceptibility $\chi(T,\alpha,T_{SP})$ is modeled in chains with variable frustration $\alpha$. 
The possible realization of BOW systems is discussed qualitatively using the different $T$ dependencies of $\delta(T,\alpha)$ and $\delta_e(T,\alpha)$. 
Section~\ref{sec5} is brief summary.

\section{\label{sec2}$J_1-J_2$ Model and SP transition}
We consider chains of $N = 4n$ spins with periodic boundary conditions and set $J_1 = 1$ as the unit of energy. We seek the $N \to \infty $ limit of the model Hamiltonian
\begin{eqnarray}
	H(\delta,\alpha;T_{SP}) &= N^{-1}  \sum_{r=1}^N  \left( 1-\delta (-1)^r  \right)   \vec{S}_r \cdot \vec{S}_{r+1} \nonumber \quad   \\
	& + N^{-1} \sum_{r=1}^N \alpha \vec{S}_{r} \cdot \vec{S}_{r+2} 
	+ \frac {\delta^2} {2\varepsilon_d}.   \qquad \quad
	\label{eq:j1j2_dim}
\end{eqnarray}
The thermodynamics at \textit{constant} $\delta$ is governed by $J_1$ and $\alpha$. There are several notable $\delta = 0$ special cases: $\alpha = 0$ 
is the HAF that has been extensively studied theoretically and widely applied to magnetic data; $\alpha_c = 0.2411$ is the quantum critical 
point~\cite{nomura1992} between the gapless and dimer phases; $\alpha = 1/2$ is the MG point~\cite{ckm69b}. The XY model has $\alpha = 0$ 
but without the $S^z_rS^z_{r+1}$ terms; it is a half-filled band of noninteracting spinless fermions. The $S^z_rS^z_{r+1}$ terms introduce nearest 
neighbor interactions between fermions.

The lattice part of Eq.~\ref{eq:j1j2_dim} is conventional~\cite{su1980,pouget2017,special_neural} with linear coupling $\delta$, harmonic potential $\delta^2/2\varepsilon_d$ 
and $T$-independent stiffness $1/\varepsilon_d$. The rigid $\delta= 0$ chain is the limit $\varepsilon_d \to 0$. The adiabatic (Born-Oppenheimer) and 
mean-field approximations ensure equal $\delta$. Minimization of the ground state energy with respect to $\delta$ gives the $T = 0$ dimerization $\delta(0,\alpha)$ 
of the chain with frustration $\alpha$. The electronic free energy per site at temperature $T$ and dimerization $\delta$ is
\begin{equation}
 A(T,\delta,\alpha)=-T N^{-1} \ln Q (T,\delta,\alpha,N).
\label{eq:freeen}
\end{equation}
The size dependence disappears with increasing $T$ or $N$ or both. The canonical partition function $Q(T,\delta,\alpha,N)$ of finite 
systems is the Boltzmann sum over the energy spectrum $E_j(\delta,\alpha,N)$ of Eq.~\ref{eq:j1j2_dim}. 

The total free energy is 
minimized to obtain the equilibrium dimerization $\delta(T,\alpha)$
\begin{equation}
      \frac{\delta(T,\alpha)}{\varepsilon_d}=-\left(\frac{\partial A(T,\delta,\alpha)}{\partial \delta} \right)_{\delta(T,\alpha)}.
\label{eq:potentialen}
\end{equation}
Since the $\delta= 0$ chain is regained for $T \ge T_{SP}$, the stiffness $1/\varepsilon_d$ is given by the curvature at 
$\delta= 0$, $-(\partial^2 A(T_{SP},\delta,\alpha)/\partial \delta^2)_0$. The electronic system sets the driving force $\partial A(T,\delta,\alpha)/\partial \delta$ 
at frustration $\alpha$. The stiffness $1/\varepsilon_d$ is the model parameter that governs $T_{SP}$, or \textit{vice versa}.

Eq.~\ref{eq:potentialen} directly gives $\delta(0)$ and $T_{SP}$ for free fermions with gap $4\delta(0)$ between the filled valence and empty 
conductions bands. However, the thermodynamic limit of the free energy of interacting fermions is not known in general, not even for the HAF. We solve Eq.~\ref{eq:potentialen} 
at finite $N$. The thermodynamic limit is reached at high $T$ when $N^{-1}lnQ(T,\delta,\alpha,N)$ becomes size independent. High-$T$ results are 
obtained~\cite{sudip19} by exact diagonalization (ED) of Eq.~\ref{eq:kekule} up to $N = 24$ for $\delta= 0$ or up to $N = 20$ for $\delta > 0$. 
DMRG calculations yield the low-energy states $E_p(\delta,\alpha,N)$ for larger $N$. The 
thermodynamic limit holds~\cite{sudip19} for $T > T^\prime(\alpha,N)$, where $T^\prime$ is the maximum of $S_C(T,\alpha,N)/T$ and $S_C$ 
is a truncated entropy with an energy cutoff. Convergence with system size can be followed~\cite{sudip19,sudipsp2020} directly and checked against 
other numerical methods or exact results. Quantitative modeling of SP transitions requires $T_{SP}(\alpha) > T^\prime(\alpha,N)$ 
for the largest $N$ considered. Convergence at $T < T_{SP}(\alpha)$ is ensured by the increasing gap on dimerization.

\section{\label{sec3}Bond-bond correlation functions } 

In this Section we consider spin correlations of the $J_1-J_2$ model, Eq.~\ref{eq:j1j2_dim}, with $\delta = 0$, 
frustration $\alpha$ and periodic boundary conditions. The ground state $\vert G,\alpha \rangle$ is nondegenerate in the gapless 
phase $\alpha \le \alpha_c$. Since the degenerate ground states of the gapped phase are even and odd under inversion at sites, 
the symmetry adapted $\vert G,\alpha,\pm 1 \rangle$  are also nondegenerate. Correlations between bonds $r$, $r + 1$ and $r^\prime$, $r^\prime + 1$ 
then depend only on $p = r^\prime - r$ and are described by four-spin correlation functions
\begin{equation}
	 C_4(p,\alpha)=\langle S^z_r S^z_{r+1} S^z_{r+p} S^z_{r+p+1}  \rangle.
\label{eq:4spincor}
\end{equation}
At the MG point, we evaluate $C_4(p,1/2)$ for $ (\vert K1 \rangle \pm \vert K2 \rangle)/\sqrt{2}$. 
Bonds separated by odd $p$ are in different diagrams and return $C_4(p,1/2) = 0$. When $p$ is even, $C_4(p,1/2)$ 
is $1/16$ or $0$, respectively, for the diagram that contains both or neither bond. It follows that $32 C_4(p,1/2) = 1$ or $0$ for even or odd $p$.

Spin-spin correlation functions also depend only on $p = r^\prime - r$ in systems with periodic boundary conditions,
\begin{equation}
	  C_2(p,\alpha)=\langle S^z_r S^z_{r+p} \rangle.
\label{eq:2spincor}
\end{equation}
Spin-spin correlations are critical in the gapless phase $\alpha \le \alpha_c$ and have been extensively characterized~\cite{sandvik2010,affleck89} for 
the HAF where $C_2(p,0)$ goes as $(-1)^p(\ln p)^{1/2}/p$ for $p \gg 1$. The range is finite in gapped phases and is just nearest 
neighbors at the MG point. When distant bonds are uncorrelated, four-spin correlation functions factor as
\begin{equation}
 C_4(p,\alpha)=C_2(1,\alpha)^2. \qquad \qquad (p \gg 1)
\label{eq:4_2_cor_reln}
\end{equation}
Since the exact $C_2(1,0)$ is $-(\ln 2 - 1/4)/3$, the $p \to \infty$ limit is $32 C_4(p,0) = 0.698$ and is approached from below as $p^{-2}$.

We anticipate that bond-bond correlations are long ranged in gapped phases with electronic dimerization $\delta_e(\alpha)$. We combine Eq.~\ref{eq:4_2_cor_reln} 
and $\delta_e(1/2) = 1$ for Kekul\'e diagrams to obtain correlations between distant bonds at any frustration
\begin{equation}
{C}_{4} \left(p,\alpha \right )=C_2(1,\alpha)^2 + {{\left (-1 \right )} ^ {p} {\delta}_{e} \left (\alpha \right )} / {64}. \quad (p \gg 1) 
\label{eq:4_2_cor_reln_mod}
\end{equation}
The difference between even and odd $p$, if any, is $\delta_e(\alpha)/32$.

We turn to the numerical analysis of finite chains of $N = 4n$ spins. Except at the MG point, the ground state is nondegenerate and there is 
small finite-size gap $E_\sigma(\alpha,N)$ to the singlet with the opposite inversion symmetry. We compute the correlation function of the most distant 
bonds, $p = 2n$, at frustration $\alpha$ in increasingly large $N = 4n$ chains 
\begin{equation}
 C_4(2n,\alpha)=\langle G(\alpha,4n) \vert S^z_1 S^z_2 S^z_{2n+1} S^z_{2n+2} \vert G(\alpha,4n) \rangle.
\label{eq:4spincor1}
\end{equation}
The next most distant bond has $p = 2n - 1$ and correlation function $C_4(2n - 1,\alpha)$. The difference is
\begin{equation}
 D_4(2n,\alpha)=C_4(2n,\alpha)-C_4(2n-1,\alpha). \qquad \quad
\label{eq:4spincor_diff}
\end{equation}
The size dependence is strong in the gapless phase and weak for $\alpha > 1/2$ due to large $\Delta(\alpha)$. Accordingly, we compute correlations 
to $N = 96$ for $\alpha= 0.35$ or less and to $N = 64$ otherwise.

\begin{figure}[t]
         \begin{center} \includegraphics[width=\columnwidth]{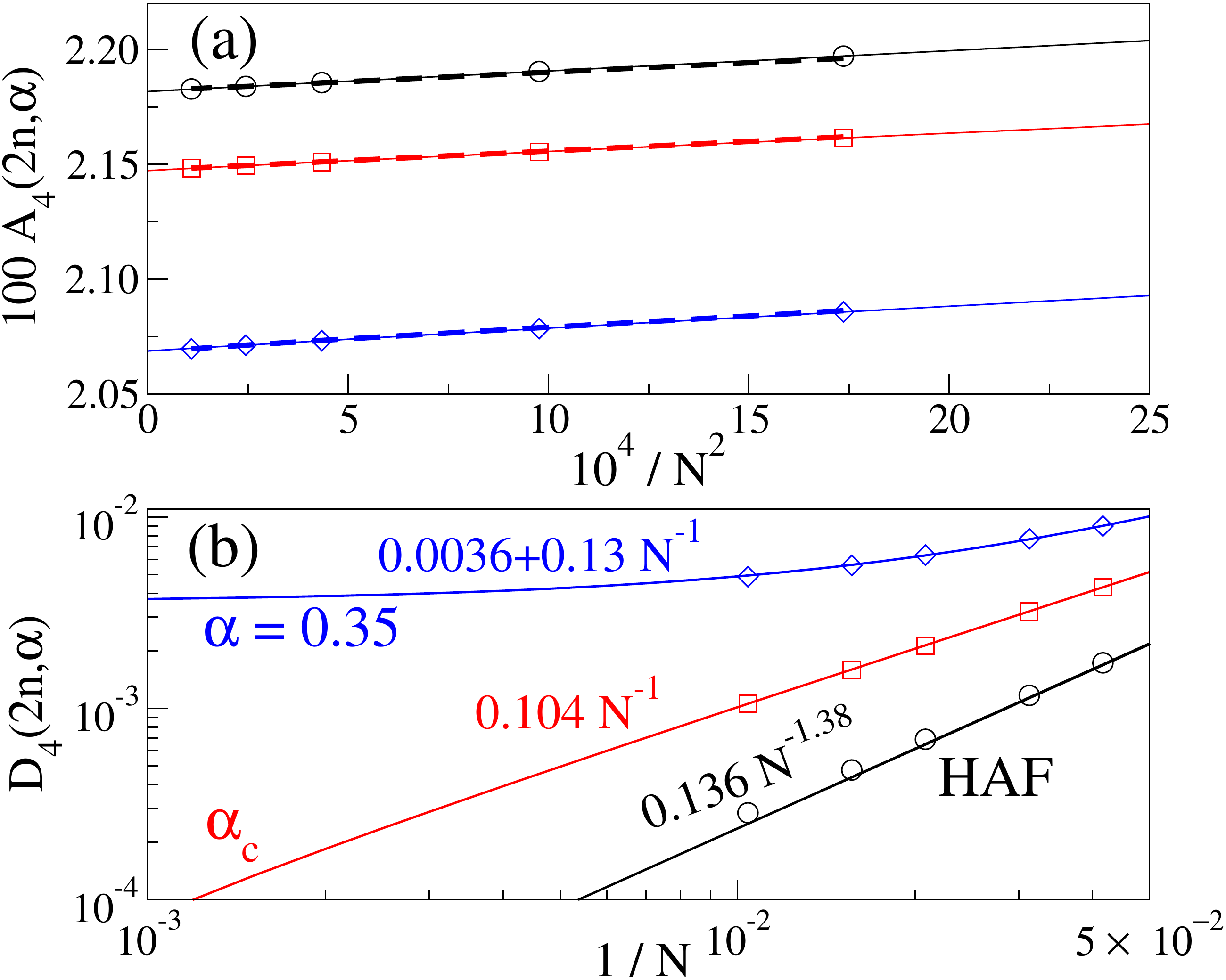}
		 \caption{Size dependence of bond-bond correlation functions $C_4(2n,\alpha)$ in Eq.~\ref{eq:4spincor1} up to $N = 4n = 96$ 
		 at $\alpha= 0$, $a_c$ and $0.35$. (a) Lines $A_4(2n,\alpha)$ are the arithmetic mean of $C_4(2n,\alpha)$ and $C_4(2n-1,\alpha)$; 
		 dashed lines are $C_2(1,\alpha,4n)^2$, Eq.~\ref{eq:4_2_cor_reln}, the square of the first neighbor spin correlation function. 
		 (b) The difference $D_4(2n,\alpha)$ in Eq.~\ref{eq:4spincor_diff}.
		 }
 \label{fig2}
 \end{center}
 \end{figure}

Fig.~\ref{fig2} shows the size dependence of bond-bond correlations at $\alpha= 0$, $\alpha_c$ and $0.35$. The solid line $A_4(2n,\alpha)$ in 
the upper panel is the arithmetic mean of $C_4(2n,\alpha)$ and $C_4(2n - 1,\alpha)$ from $N = 24$ to $96$; the dashed line is $C_2(1,\alpha,4n)^2$, 
the square of the nearest neighbor spin correlation function at system size $N = 4n$. They are equal within our numerical accuracy and go as $N^{-2}$. 
The extrapolated $\alpha= 0$ intercept agrees with the exact $2.182$ based on Eq.~\ref{eq:4_2_cor_reln}. The lower panel shows $D_4(2n,\alpha)$ from $N = 24$ 
to $96$. The size dependence at $\alpha_c$ is remarkably linear in $1/N$ as expected. The decrease is faster at $\alpha = 0$, again as expected, while $\alpha = 0.35$ 
in the gapped phase has finite $D_4 = 0.0036$ in the thermodynamic limit and electronic dimerization $\delta_e(0.35)  = 0.115$ in Eq.~\ref{eq:4_2_cor_reln_mod}. 
The $D_4(2n,\alpha)$ results confirm that long-range bond-bond correlations are finite in the dimer phase.

\begin{figure}[t]
         \begin{center} \includegraphics[width=\columnwidth]{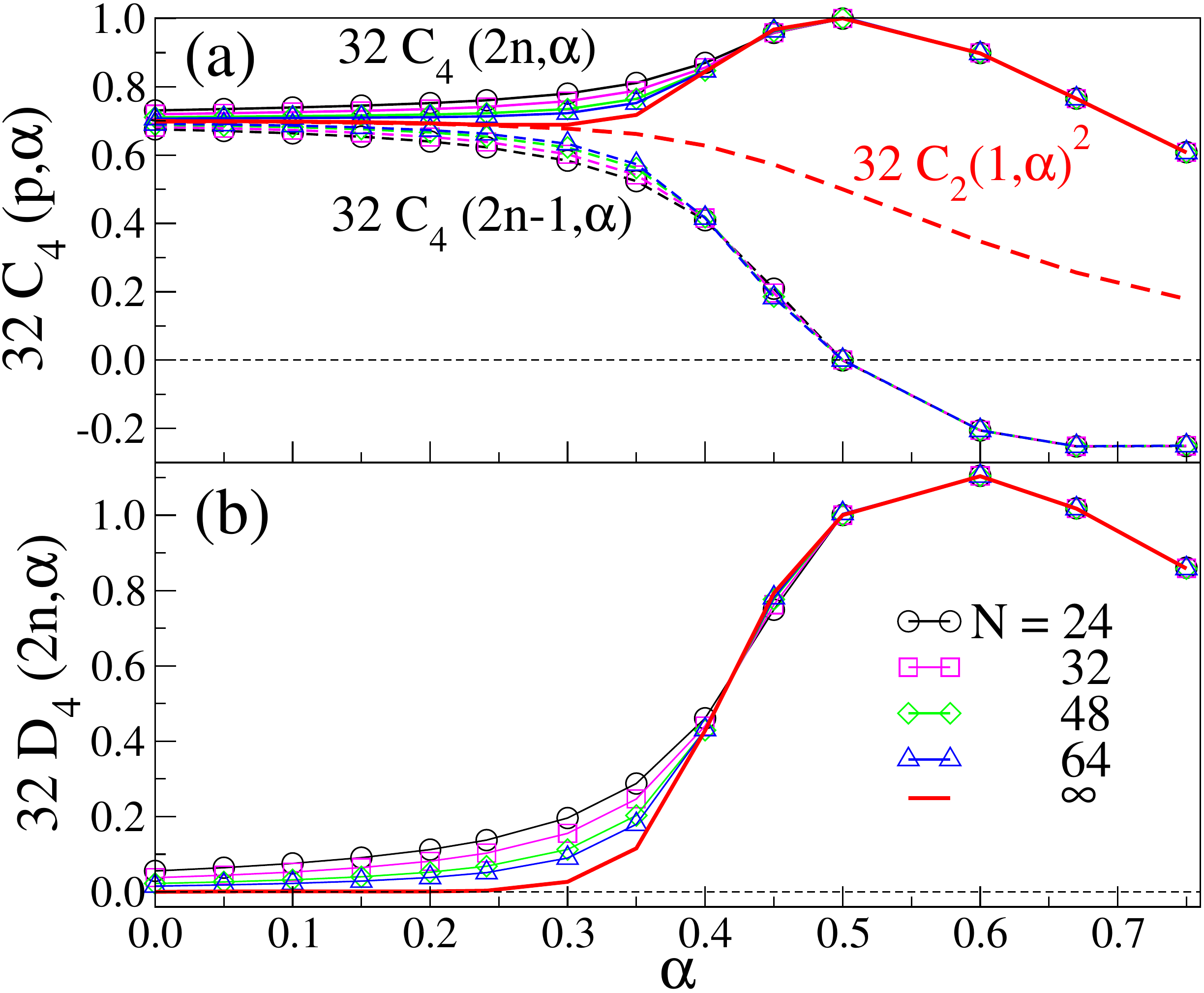}
 \caption{
	 Bond-bond correlation functions at frustration $\alpha$ for $N = 4n$ spins in Eq.~\ref{eq:j1j2_dim}. (a) $C_4(2n,\alpha)$ and 
		 $C_4(2n-1,\alpha)$ in Eq.~\ref{eq:4spincor_diff} are the most and second most distant bonds; $C_2(1,\alpha)$ is the thermodynamic limit of 
		 the first neighbor spin correlation function. (b) $32 D_4(2n,\alpha)$ is electronic dimerization $\delta_e(\alpha)$ in the limit
		 $n \to \infty.$
                 } 
 \label{fig3}    
 \end{center}
 \end{figure}

The upper panel of Fig.~\ref{fig3} shows bond-bond correlation functions at $0 \le \alpha \le 3/4$ with unit amplitude at the MG point. $C_4(2n,\alpha)$ 
correlations of bonds in the same Kekul\'e diagram decrease with system size while the $C_4(2n-1,\alpha)$ correlations of bonds in different diagrams 
increase. They converge in the gapless phase to $C_2(1,\alpha)^2$ in the thermodynamic limit. The dashed line extends $C_2(1,\alpha)^2$ 
into the gapped phase. The lower panel shows $D_4(2n,\alpha)$ for distant bonds in finite systems whose thermodynamic limit is $\delta_e(\alpha)/32$ 
in Eq.~\ref{eq:4_2_cor_reln_mod}. For $\alpha- 1/2 \ll 1$, the lowest-order corrections to $\vert K1 \rangle $ are VB diagrams in which the adjacent 
paired spins such as $(1,2)(3,4)$ are paired instead as $(1,4)(2,3)$. Adjacent pairs $(2,3)(4,5)$ in $\vert K2 \rangle$ are changed to $(2,5)(3,4)$.

Standard VB methods~\cite{ramasesha84} verify that such corrections decrease $C_4(2n,1/2)$ for large $n$ and lead to $C_4(2n-1,\alpha) < 0$ for $\alpha > 1/2$. 
The $\alpha > 1/2$ ground state $\vert G,\alpha \rangle$  that corresponds to $\vert K1 \rangle$ at $\alpha = 1/2$ has antiferromagnetic correlations 
$C_2(1,\alpha)$ between spins $2r$, $2r-1$ and ferromagnetic correlations between spins $2r$, $2r +1$. The spin correlations are reversed in 
the ground state that corresponds to $\vert K2 \rangle $ at $\alpha= 1/2$. The net result is a maximum around $\alpha \sim 0.60$ in the lower 
panel that corresponds to $\delta_e(0.6) > 1$ in one broken-symmetry ground state and $\delta_e(0.6) < -1$ in the other.

To understand the thermodynamic limit of bond-bond correlations, it is instructive to consider the $J_1-J_2$ model at $\alpha > 1$. 
The $J_1 = 0$ limit ($\alpha \to \infty$) corresponds to noninteracting HAFs on sublattices of odd and even numbered sites, 
respectively, with known spin correlation functions. It follows immediately at $J_1 = 0$ that (1) spin-spin correlation functions such as $C_2(1,0)$ 
with odd $p = r^\prime - r$ in Eq.~\ref{eq:2spincor} are identically zero and that (2) bond-bond correlation functions in Eq.~\ref{eq:4spincor} factor into 
products of sublattice spin-spin correlation functions such as
\begin{equation}
 {C}_{4} \left (2n \right ) = \left \langle {S}_{1}^{z} {S}_{2n+1}^{z} \right \rangle \left \langle {S}_{2}^{z} {S}_{2n+2}^{z} \right \rangle.  \quad  \left ({J}_{1} =0 \right ) \qquad
\label{eq:4spincor_factor}
\end{equation}
Since both HAF correlations are between $n$th neighbors, $C_4(2n)$ is positive and decreases as $n^{-2}$ for distant spins. 
The corresponding expression for $C_4(2n - 1)$ has spin correlations between $n$th neighbors of one sublattice and $(n - 1)$th neighbors on 
the other; $C_4(2n - 1)$ is negative and increases as $n^{-2}$. All correlation functions in Fig.~\ref{fig3} are zero at $J_1 = 0$ in the thermodynamic 
limit. The gapped phase terminates at the quantum critical point~\cite{soos-jpcm-2016} $\alpha_2 = 2.27$ at the onset of a gapless 
decoupled phase with nondegenerate ground state and quasi-long-range-order at wave vector $q = \pi/2$.

 \begin{figure}[t]
         \begin{center} \includegraphics[width=\columnwidth]{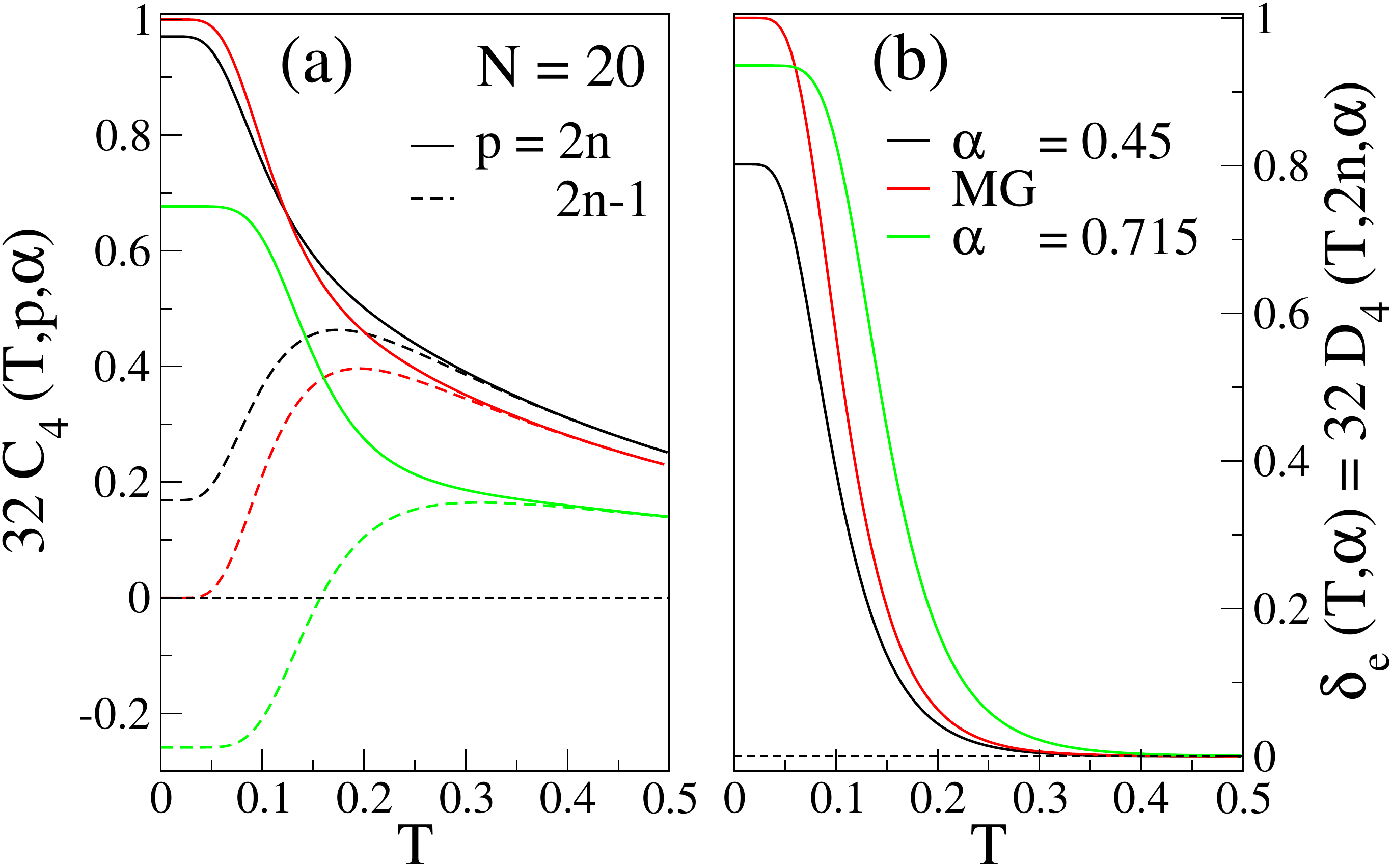}
 \caption{(a)Temperature dependence of bond-bond correlation functions $C_4(T,p,\alpha)$ at system size $N = 4n = 20$ and $\alpha = 0.45$, $0.50$ and $0.715$; 
		 lines for $p = 2n$, dashed lines for $p = 2n - 1$. (b) electronic dimerization $\delta_e(T,\alpha) = 32 D_4(T,2n,\alpha)$.
                 }
 \label{fig4}
 \end{center}
 \end{figure}

The $T$ dependence of bond-bond correlation functions is obtained as usual. The partition function $Q(T,\alpha,N)$ is the sum of 
$\exp(-\beta E_j(\alpha,N))$ over $2^{N}$ states ($\beta= 1/k_BT$). We have
%
%

%
\begin{eqnarray}
	{C}_{4} \left (T,p,\alpha,N \right ) = \frac {1} {\left (Q(T,\alpha,N) \right )} \times \qquad  \qquad \qquad \nonumber \\
        \sum_{j} \left \langle j \vert {S}_{1}^{z} {S}_{2}^{z} S_{1+p}^{z} {S}_{2+p}^{z} \vert j \right \rangle \exp(-\beta {E}_{j} \left (\alpha,N \right )) . \qquad
        \label{eq:4spincor_temp}
\end{eqnarray}
Similar expressions hold for the spin-spin correlations $C_2(T,p,\alpha,N)$. Fig.~\ref{fig4} presents ED results for $N = 20$ 
at several $\alpha$ with large $\Delta(\alpha)$. The finite size gap $E_\sigma(\alpha,N)$ contributes at $T \sim 0$ but hardly 
matters around $T \sim \Delta(\alpha)$ in the region of interest where $N = 20$ is in, or almost in, the thermodynamic limit. 
We measured excitation energies in Eq.~\ref{eq:4spincor_temp} from $E_\sigma(\alpha,N)/2$ and averaged the matrix elements of the two singlets.

$T$ initially suppresses correlations between distant bonds, as shown by increasing $C_4(T,2n - 1)$ for bonds in different Kekul\'e diagrams 
and decreasing $C_4(T,2n)$ for bonds in the same diagram. The $T$ dependence of $C_2(T,1,\alpha)^2$ is related to the gap $\Delta(\alpha)$. 
Electronic dimerization $\delta_e(T,\alpha)$ decreases rapidly when $\beta \Delta(\alpha) < 1$. 
We emphasize that all results in this section are for rigid chains with $\delta = \varepsilon_d = 0$ in Eq.~\ref{eq:j1j2_dim}.

\section{\label{sec4}Gap relations and thermodynamics}

 \begin{figure}[]
         \begin{center} \includegraphics[width=\columnwidth]{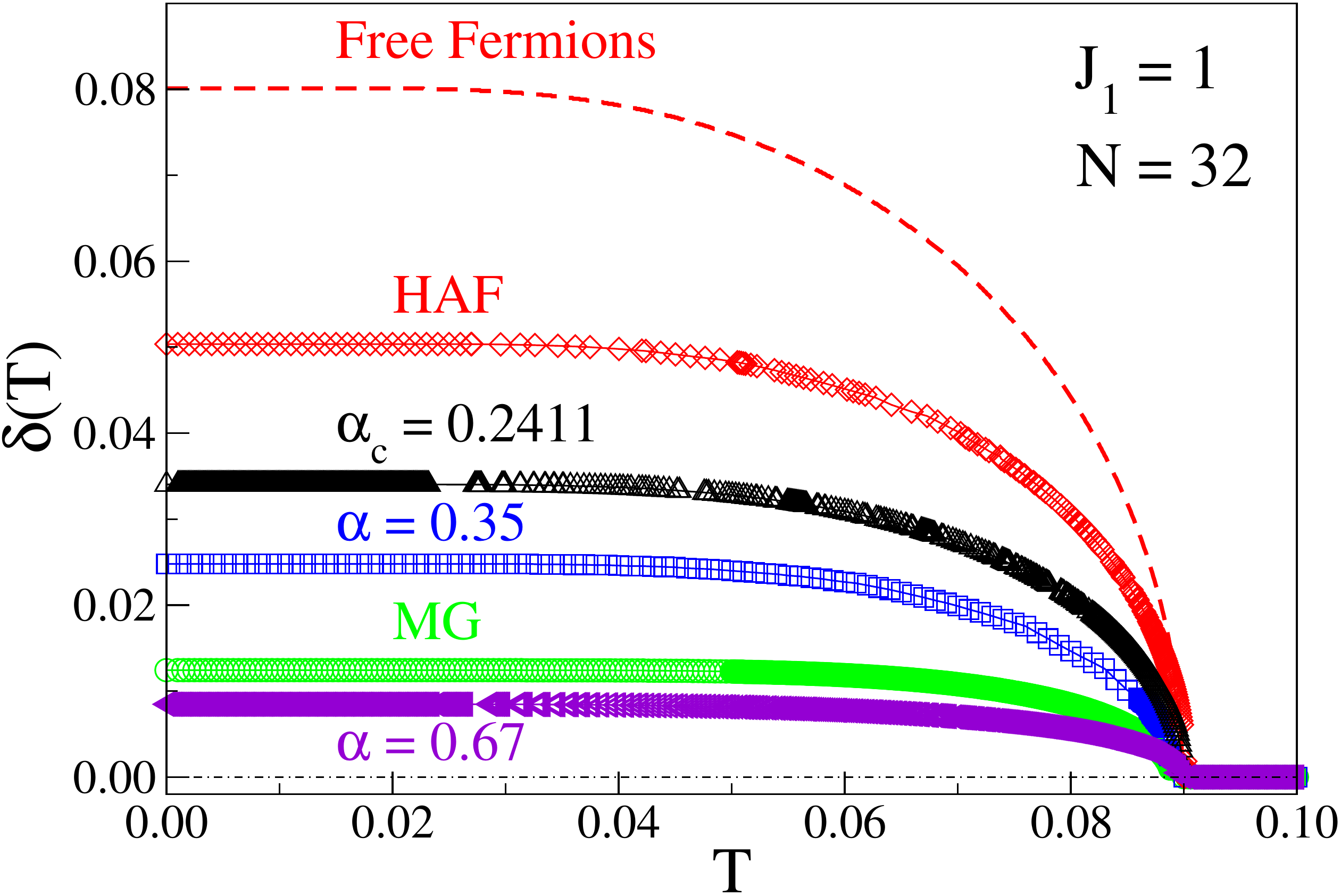}
 \caption{
	 Equilibrium dimerization $\delta(T)$ of spin chains with $T_{SP} = 0.09$ leading to stiffness $1/\varepsilon_d$ and $\delta(0)$ 
		 in Eq.~\ref{eq:potentialen}. The HAF ($\alpha = 0$), $\alpha_c = 0.2411$, $\alpha = 0.35$, MG ($\alpha= 0.50$) and $\alpha= 2/3$ 
		 curves are based on Eq.~\ref{eq:j1j2_dim} with $N = 32$ spins. Free fermions refer to a half-filled band of spinless fermions.}
 \label{fig5}
 \end{center}
 \end{figure}
The magnetic susceptibility $\chi(T)$ of the organic crystal characterized Jacobs et al.~\cite{jacob1976} followed the HAF 
quantitatively at $T > T_{SP} = 12$ K. Dimerization $\delta(T)$ opens a gap $\Delta(T)$ at the transition, and both increase 
on cooling to $T = 0$. The similarity of SP transitions to BCS superconductors was recognized from the outset~\cite{jacob1976}. 
The BCS gap equation specifies the function $\Delta(T)/\Delta(0)$ and the ratio $\Delta(0)/T_c = 3.52$. The mathematics are 
almost identical for the XY model ($\alpha = 0$, no $S_1^zS_2^z$ terms in Eq.~\ref{eq:j1j2_dim}). The $T = 0$ gap between 
the filled valence and empty conduction band of spinless fermions is $\Delta(0) = 4\delta(0)$. The equilibrium $\delta(T)$ is given 
by Eq.~\ref{eq:potentialen}, which can be solved in the thermodynamic limit for noninteracting fermions and returns $\Delta(0)/T_{SP} = 3.56$ for $T_{SP} = 0.09$ 
The same analysis holds for polyacetylene in the approximation of a half-filled tight-binding band.

There are important differences as well. The phonon-coupled attraction between free electrons ensures the same $\Delta(0)/T_c$ while the Peierls 
instability of correlated $J_1-J_2$ models leads to a range of $\Delta(0)/T_{SP}$. The ratio $T_{SP}/4J_1$ is at least 100 times larger than 
$T_c/\varepsilon_F$, where $\varepsilon_F$ is the Fermi energy; SP systems are in the regime of intermediate coupling instead of  
very weak coupling. Finally, free fermions are inappropriate for strongly correlated spin-1/2 chains.

Fig.~\ref{fig5} illustrates the dependence of the equilibrium dimerization, $\delta(T,\alpha)$ in Eq.~\ref{eq:potentialen}, on frustration 
and correlations in chains with $T_{SP}= 0.09$ ($0.09 J_1$). The free fermions curve is for a half-filled band, the XY model. The other 
curves are based on Eq.~\ref{eq:j1j2_dim} with $N = 32$ and frustration $\alpha$. Frustration clearly changes both $\delta(0,\alpha)$ and 
the $T$-dependence.

The ground-state energy per site $\varepsilon_0(\delta,\alpha)$ and singlet-triplet gap $\Delta(\delta,\alpha)$ of Eq.~\ref{eq:j1j2_dim} 
in the thermodynamic limit has been extensively discussed~\cite{barnes99,cross79} and sometimes debated, especially at $\alpha= 0$ (HAF) and $\alpha_c$, 
using field theory and numerical methods. The $\delta \ll 1$ behavior of $\varepsilon_0(\delta,\alpha)$ and $\Delta(\delta,\alpha)$ 
has been sought. HAF results and references are collected in Ref. \cite{johnston2000}; $\alpha_c$ results in Ref. \cite{mkumar2007}. Conventional DMRG 
calculations~\cite{sudipsp2020} to $N = 96$ return accurate energies $\varepsilon_0(\delta,\alpha)$ and gaps $\Delta(\delta,\alpha)$ for $\delta$ between 
$0.001$ and $0.1$.

The derivative $\varepsilon_0^{\prime}(\delta,\alpha)$ with respect to $\delta$ is zero at $\delta = 0$ in the gapless phase $\alpha \le \alpha_c$ 
and has a cusp $\vert \delta \vert$  for $\alpha > \alpha_c$. The curvature $\varepsilon_0^{\prime \prime}(\delta,\alpha)$ 
diverges at $\delta = 0$ for $\alpha \le \alpha_c$. The SP instability also varies within the gapless and gapped phases. 
The equilibrium gap is $\Delta(T,\alpha) = \Delta(\delta(T),\alpha)$. Gap relations $\Delta(0)/T_{SP}$ are listed in Table~\ref{tab1} 
for chains with $T_{SP} = 0.09$ in Fig.~\ref{fig5} along with $\delta(0)$, the stiffness $1/\varepsilon_d$ and $\Delta(\delta(0),\alpha)$. 
The free fermions relation is very close to BCS.

\begin{table}

\caption{\label{tab1}  Gap relation $\Delta(0)/T_{SP}$ of spin-$1/2$ chains with $T_{SP} = 0.09$, frustration $\alpha$ in Eq.~\ref{eq:j1j2_dim},
        stiffness $1/\varepsilon_d$, equilibrium dimerization $\delta(0)$ at $T = 0$ given by Eq.~\ref{eq:potentialen}, and
	$T = 0$ gap $\Delta(0) = \Delta(\delta(0),\alpha)$.}	
\begin{ruledtabular}
\begin{tabular}{ c  c  c  c  c }
Model &  $ \frac {\Delta(0)} {T_{SP}}$ &  $\delta(0)$  &   $\frac {1} {\varepsilon_d}$  &  $\Delta(\delta(0),\alpha)$    \\ \hline

Free fermions          &  3.56    &  0.0801  &   1.86 &  0.320$^{\rm a}$  \\
$\alpha = 0$ (HAF)     &   2.52   & 0.0504   &  2.94  &  0.227              \\
$\alpha_c = 0.2411$    &  2.54    &  0.0340  &  6.40  &  0.228            \\
$\alpha = 0.35$        &  2.65    &  0.0248  &  11.1  &  0.239            \\
$\alpha=0.50$ (MG)     &  3.88    & 0.0124   & 30.4   &  0.349               \\
$\alpha=0.67$          &  5.94    & 0.00848  &  45.0  &  0.535     \\     
\end{tabular}
$^{\rm a}$ $\Delta(0)=4\delta(0)$, half-filled band.
\end{ruledtabular}
\end{table}

We find that $\Delta(0)/T_{SP}$ increases slowly with frustration in the gapless phase with divergent curvature $\varepsilon^{\prime \prime}$. 
The modest increase at $\alpha= 0.35$ in the dimer phase reflects small $\Delta(0.35) = 0.0053$ compared to either $\Delta(\delta(0),0.35)$ or thermal 
energies at $T_{SP} = 0.09$. The MG point illustrates the decisive role of an energy cusp for degenerate ground states. Even larger $\Delta(0)/T_{SP}$ 
at $\alpha= 2/3$ is due to the larger $\Delta(2/3) = 0.433$ that increases the relative weight of the ground state at low $T$. The largest $\Delta(\alpha)$ 
at $\alpha= 0.715$ is a few percent greater.

The $\Delta(0)/T_{SP}$ in Table~\ref{tab1} are at constant $T_{SP} = 0.09$. Gap relations at intermediate coupling depend weakly on $T_{SP}$. 
For example, the HAF ratio at stronger coupling $T_{SP} = 0.15$ is $2.41$, almost $5\%$ less than in $2.52$ in Table~\ref{tab1}. Conversely, 
weaker coupling $T_{SP} = 0.06$ increases the ratio slightly and is numerically more demanding since the thermodynamic limit is 
reached at larger $N \sim 100$.

\begin{figure}[t]
         \begin{center} \includegraphics[width=\columnwidth]{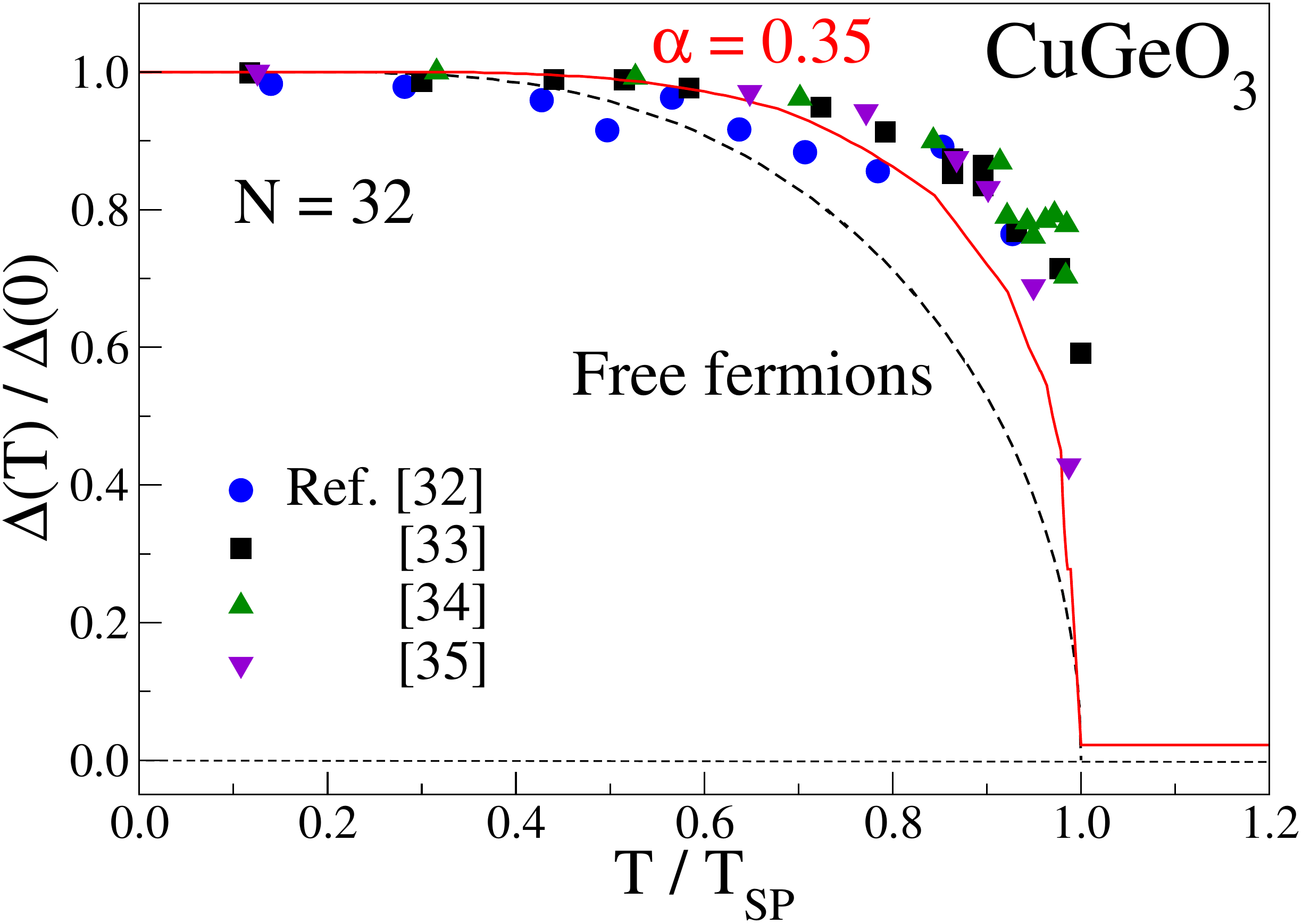}
		 \caption{Scaled singlet-triplet gap $\Delta(T)/\Delta(0)$ vs. $T/T_{SP}$. The solid red line is $\Delta(\delta(T),\alpha)/\Delta(\delta(0),\alpha)$ 
		 with $\alpha= 0.35$ in Eq.~\ref{eq:j1j2_dim}; the dashed line is $\delta(T)/\delta(0)$ for free fermions in Fig.~\ref{fig5}. 
		 The symbols are inelastic neutron data on CuGeO$_3$ crystals from Refs. \cite{nishi1994,lussier1996,martin1996,regnault96}. 
                 }
 \label{fig6}
 \end{center}
 \end{figure}

Inelastic neutron scattering (INS) from the singlet ground state is exclusively to triplets in models with isotropic exchange~\cite{mkumarbndord2010}. 
INS data on several CuGeO$_3$ crystals have been reported~\cite{nishi1994,lussier1996,martin1996,regnault96} as $\Delta(T)/\Delta(0)$ vs. $T/T_{SP}$. 
The transition temperatures 
and singlet-triplet gaps vary by $5\%$ and $10 \%$, respectively. Deviations from free fermions or BCS were emphasized in every case. 
The data are shown in Fig.~\ref{fig6} together with the gap ratios $\Delta(T)/\Delta(0)$ at $\alpha= 0.35$ (solid line) and free fermions 
(dashed line). Deviations were unexpected because the organic SP crystal, unsuitably small~\cite{regnault96} for INS, was thought~\cite{bray1983} to follow 
BCS. Correlated states with $\alpha= 0.35$ are consistent with INS, and how quantitatively remains to be seen.

\begin{figure}[t]
         \begin{center} \includegraphics[width=\columnwidth]{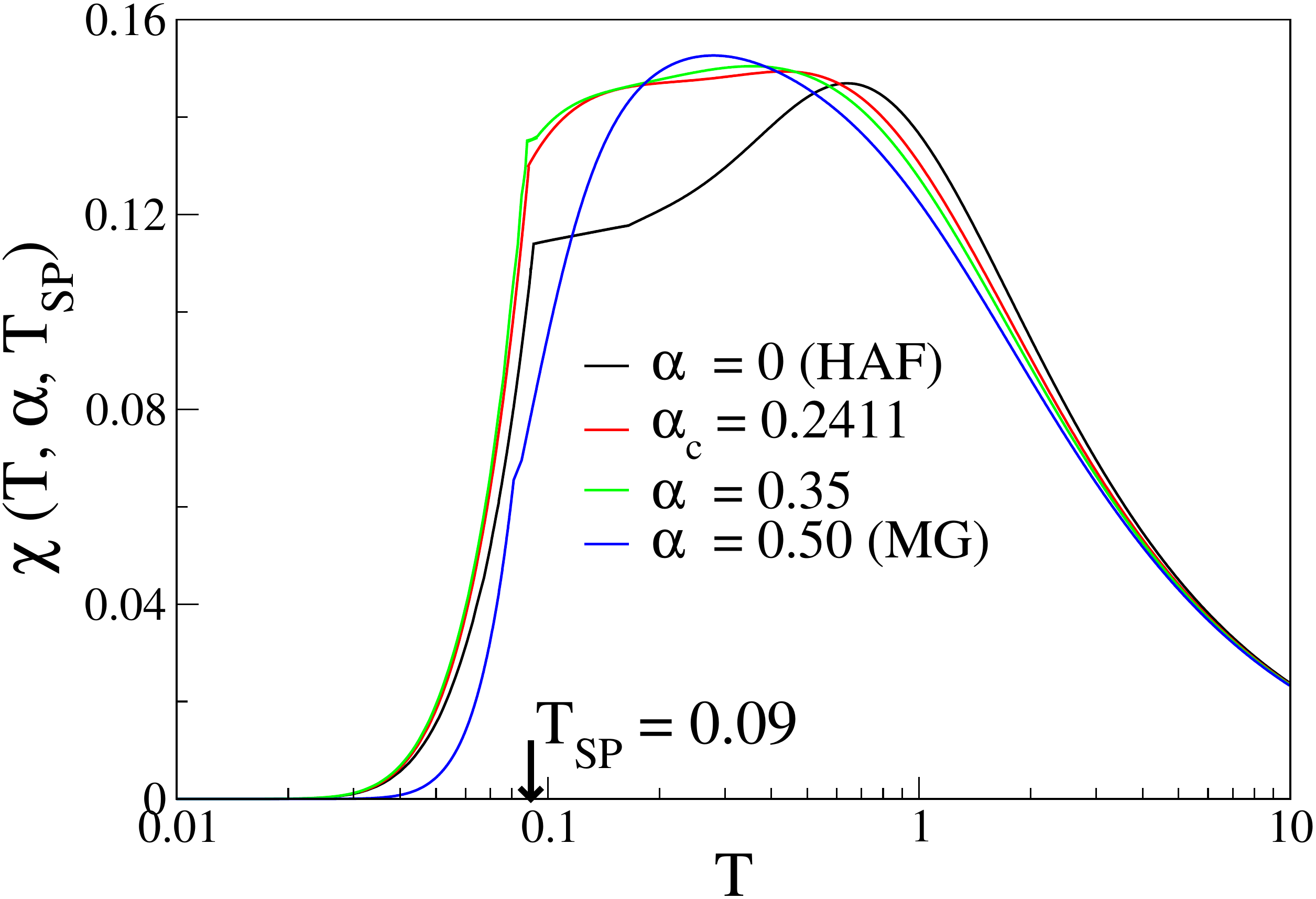}
 \caption{
	 Magnetic susceptibility $\chi(T,\alpha,T_{SP})$ of chains with $T_{SP} = 0.09$ and variable $\alpha$ in Eq.~\ref{eq:j1j2_dim}.
                 }
 \label{fig7}
 \end{center}
 \end{figure}

The thermodynamic limit holds for $T > T_{SP} = 0.09$ in $J_1-J_2$ models with $N \sim 50$ and $0 \le \alpha \le 3/4$ in Eq.~\ref{eq:j1j2_dim}. 
The calculated $\chi(T,\alpha,T_{SP})$ in Fig.~\ref{fig7} are quantitative for models with equilibrium $\delta(T,\alpha)$ given by Eq.~\ref{eq:potentialen}. 
The logarithmic scale emphasizes low $T$. The units of $\chi$ are $J_1/N_Ag^2 \mu_B^2$ where $N_A$ is Avogadro's number, $g = 2.003$ is 
the free electron value (isotropic exchange excludes spin-orbit coupling) and $\mu_B$ is the Bohr magneton. The $T > T_{SP}$ susceptibility varies 
with frustration $\alpha$ in both gapped and gapless phases, but the $\chi(T,\alpha)$ maximum around $0.14$ is roughly constant. The Curie law at high $T$ is $1/4T$.

The measured $\chi(T)$ of CuGeO$_3$ clearly indicated~\cite{riera1995,fabricius1998} $\alpha = 0.35$ and $J_1 = 160$ K in crystals with $T_{SP} = 14.4$ K. Electron 
spin resonance (ESR) with the applied magnetic field along the crystal $c$ axis fixed~\cite{fabricius1998} $g = 2.256$. The parameters $\alpha$, $J_1$ and $T_{SP}$ 
account for $\chi(T)$ over the entire range to $950$ K and for the specific heat anomaly, as shown in Figs. 5 and 6 of Ref. \cite{sudipsp2020}. 
The coupling $T_{SP}/J_1 = 0.09$ motivated the choice of $T_{SP}$ in Fig.~\ref{fig5}. The measured~\cite{jacob1976} $\chi(T)$ of the organic SP 
crystal TTF-CuS$_4$C$_4$(CF$_3$)$_4$ with $T_{SP} = 12$ K and $g = 1.997$ from ESR is quantitatively modeled (Fig. 5, Ref. \cite{sudipsp2020}) with $J_1 = 79$ K, $\alpha= 0$ 
and $T_{SP}$. Since the coupling $T_{SP}/J_1 = 0.15$ is stronger, the thermodynamic limit is reached at $N \sim 32$.

The initial $\chi(T)$ analysis~\cite{jacob1976} was based on the HAF with correlated states for $T > T_{SP}$ and a mean-field approximation at 
lower $T$. As noted explicitly, the quantitative fit shown in Fig. 5 of Ref. \cite{jacob1976} or in {Fig. 10 of Ref. \cite{bray1983}} 
required another parameter. The free fermions or BCS relation with $T_{SP} = 12$ K gave the observed $T$ dependence but overestimated $\Delta(0)$ while the 
parameter $\Delta(0)$ needed for $\chi(T)$ at low $T$ returned~\cite{jacob1976} $T_{SP} = 9$ K, a $25\%$ discrepancy. The inconsistency is due to the 
mean-field approximation. The influential but incorrect conclusion~\cite{bray1983} that the HAF gap relation is close to BCS strongly influenced the discussion 
of INS data in Fig.~\ref{fig6}.

The $J_1-J_2$ model with $\delta= 0$, $\alpha \ne 0$ in Eq.~\ref{eq:j1j2_dim} has many additional quantum phases~\cite{hikihara2008,sudan2009} 
when $J_1 < 0$ and an applied magnetic field are considered as well as $J_1 > 0$. Physical realizations of gapped phases are very rare, 
however. CuGeO$_3$ may well be the first. Half-filled Hubbard-type models also support gapped quantum phases~\cite{nakamura2000,sengupta2002,mkumar2009} with doubly 
degenerate BOW ground states and broken inversion symmetry at sites, but only over narrow ranges of parameters. Such models have charge 
as well as spin degrees freedom. Dimerization is associated with alternating transfer integrals $t(1 \pm \delta)$ along the chain. BOW 
phases and thermodynamics are less well characterized, both because interactions between charges require effective parameters that 
are poorly understood and because the models have about $4^{N}$ instead of $2^{N}$ states. 

The defining features of BOW phases are magnetic and spectroscopic evidence for broken inversion symmetry at atomic or molecular 
sites in crystals whose structure has inversion symmetry ($\delta = 0$) at sites. Inversion symmetry at sites is trivially broken 
on dimerization ($\delta \ne 0)$. It is also broken in linear combinations of the degenerate ground states $\vert G,\alpha,\pm 1 \rangle$ 
of $\delta= 0$ chains with $\alpha > \alpha_c$ and $\delta_e(\alpha) \ne 0$.

\begin{figure}[t]
         \begin{center} \includegraphics[width=\columnwidth]{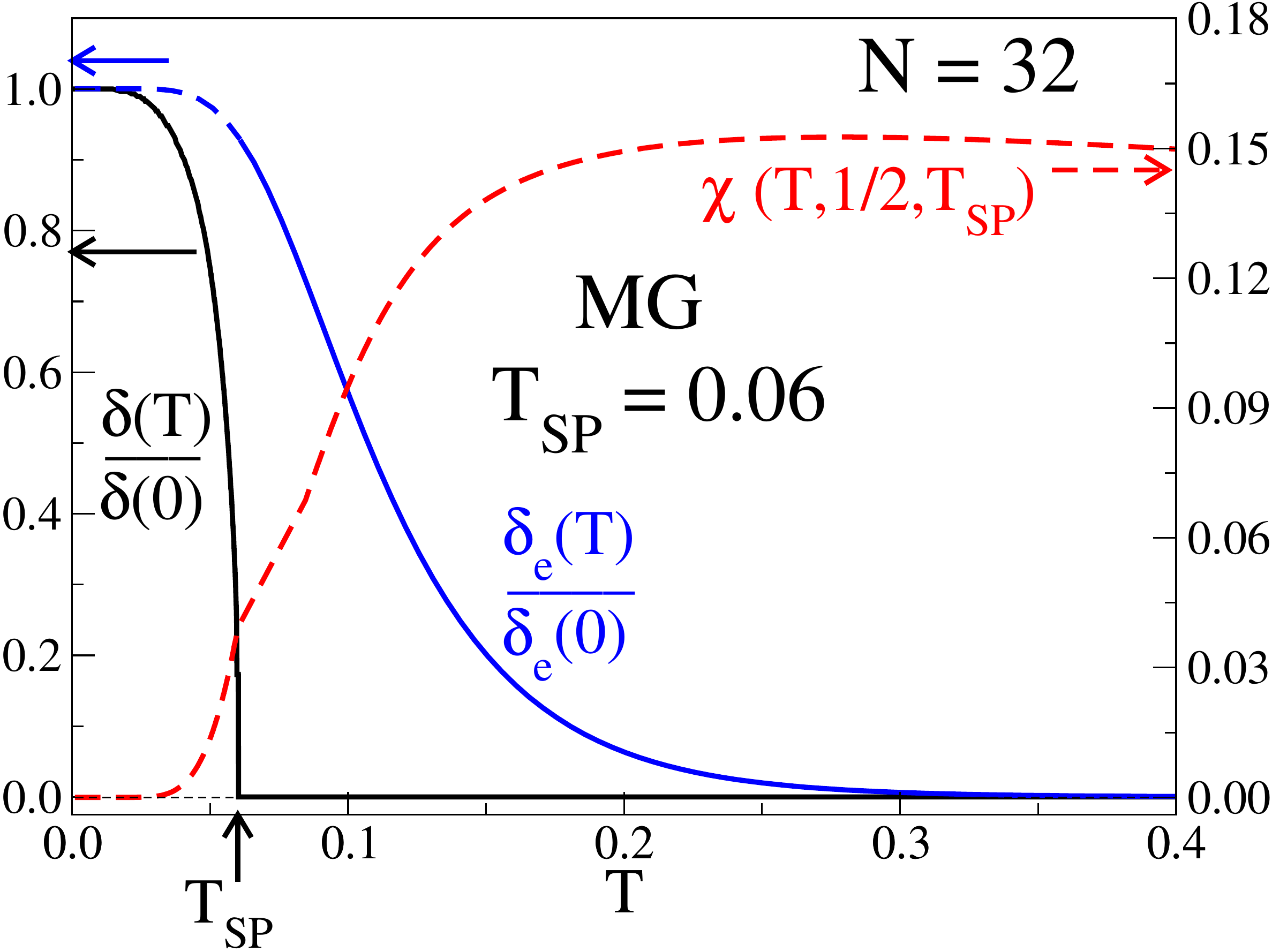}
		 \caption{Equilibrium dimerization $\delta(T,1/2)$ and susceptibility $\chi(T,1/2,T_{SP})$ at the MG point for chains 
		 with $T_{SP} = 0.06$ and $N = 32$ spins. Electronic dimerization $\delta_e(T,1/2)$ for $T \ge T_{SP}$ is discussed together 
		 with Eq.~\ref{eq:4spincor_temp}. The finite chain with $\alpha = 1/2$ and $T_{SP} = 0.06$ is in the thermodynamic limit. 
                 }
 \label{fig8}
 \end{center}
 \end{figure}

The analysis at the MG point holds qualitatively in general. We contrast in Fig.~\ref{fig8} the $T$ dependencies of \textit{structural} 
dimerization $\delta(T)/\delta(0)$ and \textit{electronic} dimerization $\delta_e(T)/\delta_e(0)$ in chains with ${\alpha = 1/2}$ and ${T_{SP} = 0.06}$. The magnitude of $\chi(T,1/2,T_{SP})$
is given on the right (note the different scale). The gap $\Delta(1/2) = 0.233$ reduces the susceptibility maximum 
around $T \sim 0.25$ by over $50 \%$ at $T_{SP} = 0.09$ and by even more at ${T_{SP} = 0.06}$. The stiffness $1/\varepsilon_d$ is the parameter related to 
$T_{SP}$ in deformable chains subject to Peierls or SP transitions and equilibrium
$\delta(T,\alpha$) at $T \le T_{SP}$ while long-range bond-bond correlations leads to 
$\delta_e(T)$ in gapped deformable chains for $T \ge T_{SP}$.

In the present context, either dimerization can be studied by an appropriate probe
of the electronic wave function, for example an intensity $I(T)$. Hubbard-type models 
have long been applied to face-to-face stacks of planar $S = 1/2$ molecular ions
in organic salts or charge-transfer crystals~\cite{soos1974}. Regular ($\delta = 0$) stacks with 
inversion symmetry at sites have Raman active, IR silent, totally symmetric ($a_{g}$) vibrations. Some $a_{g}$ modes become strongly IR active in dimerized stacks, as also
seen in polyacetylene, and are analyzed in terms of electron-phonon coupling~\cite{rice1979,girlandosoos2004}. Since
the IR intensity $I(T)$ turns on sharply at the transition, it provided a convenient 
and widely used way to identify Peierls transitions.

The symmetry argument also applies to BOW phases that were not recognized at the time 
and where $I(T)$ follows $\delta_e(T)$ for $T \ge T_{SP}$. Smoothly decreasing $I(T)$ were reported~\cite{cesare1977,girlando1984} as 
in Fig~\ref{fig8} in rare exceptional cases. The estimated $T_P \sim 220$ K of Rb-TCNQ(II) 
crystals was based~\cite{cesare1977} on $I(T_P)/I(25) \sim 1/2 \sim \chi(T_P)/\chi(300)$. The $100$ K crystal structure~\cite{soosqueen09} was 
subsequently found to be in the same space group, triclinic
$P \overline{1}$, as the 
$295$ K structure, thereby definitively ruling out a transition between $100$ 
and $300$ K. The evidence for a BOW system is compelling but qualitative. The $300$ K structure of TMPD-TCNQ crystals has regular ($\delta = 0$) 
stacks and a large singlet-triplet gap. The IR intensity $I(T)$ decreased~\cite{girlando1984} gradually with $T$ to $I(300)/I(15) \sim 20 \%$ and to $10 \%$ at 350 K, the limit 
of thermal stability. The structure at lower $T$ has not been reported. The peculiar combination 
in BOW systems of sites with $C_i$ symmetry, large magnetic gaps and broken electronic $C_i$ symmetry occurs 
naturally in gapped phases with doubly degenerate ground states.


\section{\label{sec5}Discussion}

We have discussed the SP transition of the $J_1-J_2$ model, Eq.~\ref{eq:j1j2_dim}, with first neighbor exchange $J_1$, variable frustration $\alpha= J_2/J_1$ 
and linear coupling to a harmonic lattice. The thermodynamic limit of correlated states is reached at finite system size that depends on $T_{SP}/J_1$ 
or on $\Delta(0)$ in gapped rigid chains. The ED/DMRG procedure requires a microscopic model for $\delta = 0$ chains since the ground state instability drives the transition. In that case, the results are model exact for gap relations or bond-bond 
correlations or thermodynamics. However, microscopic 1D models with $\delta= 0$ chains are inevitably approximations to quasi-1D systems.

We quantified long-range bond order in the degenerate ground states $\vert G, \alpha, \pm 1 \rangle$ of gapped phases using the four-spin 
correlation functions $C_4(p,\alpha)$ in Eq.~\ref{eq:4spincor}. Linear combinations of  $\vert G, \alpha, \pm 1 \rangle$ are 
BOWs with broken inversion symmetry and electronic dimerization $\delta_e(0,\alpha)$ that remains finite for $\Delta(\alpha)/T > 1$ 
in models with inversion symmetry at sites. The different $T$ dependencies of $\delta(T,\alpha)$ for $T \le T_{SP}$ 
and $\delta_e(T,\alpha)$ could in principle be evidence for physical realizations of BOW systems.

Two comments are in order to place the present results in broader context. First, there is far more to Peierls systems than a transition; 
that includes all structural, spectroscopic, transport and thermodynamic characterization of condensed matter systems. The comprehensive 
review of Heeger et al.~\cite{heeger88} of trans-polyacetylene, a prototypical Peierls system, discusses the consequences of dimerization, 
but there is no transition up to the limit of thermal stability.  Some planar $\pi$-electron donors (D) and acceptors (A) 
crystallize in mixed face-to-face stacks $...D^\rho A^{-\rho} D^\rho A^{-\rho}...$  with charge transfer $\rho$ that varies from
$\rho \sim  0$ in neutral closed-shell molecules to $\rho \sim 1$ spin-1/2 radical ions. The reviews in Ref.~\cite{special_neural} describe 
their neutral-ionic transitions or crossovers on cooling or under pressure. Inversion symmetry at sites at high $T$ is broken on dimerization. 
Microscopic modeling becomes far more challenging in correlated systems with both charge and spin degrees of freedom. Refs. \cite{jerome2004} and \cite{pouget2017}
discuss the characterization of other quasi-1D systems with charge and spin degrees of freedom and Peierls transitions to incommensurate as well as dimerized chains.

Second, the observed structural changes at Peierls transitions are 3D rather than restricted to 1D chains or stacks. Conversely, a dimerization 
transition may have other origins. The cation radical stack in Wurster's blue perchlorate dimerizes at $180$ K when perchlorate 
rotation stops. Considerable evidence and analysis are needed to recognize crystals with quasi-1D spin chains and a 
transition driven by a Peierls instability. Weak interactions between chains are more likely to induce a 2D or 3D transition at low $T$.

The magnitude of $T_{SP}/J_1$ facilitated modeling~\cite{sudipsp2020} the SP transitions of TTF-CuS$_4$C$_4$(CF$_3$)$_4$ and CuGeO$_3$. 
The numerical problem is far greater at, say, $T_{SP}/J_1= 0.01$ because the thermodynamic limit at $\delta= 0$ would be reached at much larger $N$, 
the chain would be stiffer and the $T = 0$ dimerization $\delta(0)$ smaller. The striking difference between the gapless phase at $\alpha_c$ 
and the gapped phase at $\alpha = 0.35$ phase, which is not evident in Table~\ref{tab1}, would be manifest. The gap $\Delta(0.35) = 0.0053$ 
that hardly matters for thermodynamics at $T \sim 0.09$ could not be ignored at $T \sim 0.01$. Small $T_{SP}/J_1$ brings out 
the qualitative difference between a nondegenerate ground state with divergent $\varepsilon_0^{\prime \prime}(\delta)$ at $\delta= 0$ 
and a doubly degenerate ground state with discontinuous $\varepsilon_0^{\prime}(\delta)$ at $\delta= 0$. The difference is evident at 
the MG point when $T_{SP}/J_1 = 0.09$.

Isotropic exchange is assumed in 1D models studied by field theory or numerical methods, and that is indeed the dominant 
magnetic interaction. But the models are approximate because corrections to isotropic exchange due to spin-orbit coupling are neglected. 
They are incomplete because dipolar interactions between spins are neglected, as are hyperfine interactions with nuclear spins and 
interactions of any kind between chains. More complete specific models are required for the low-$T$ properties of materials with quasi-1D 
spin chains. Simpler 1D models with isotropic exchange are adequate for SP transitions and thermodynamics.


\section*{Acknowledgments} 
MK did majority of this work at GRTA IIT(BHU). 
MK thanks DST India for financial support through a Ramanujan fellowship.
ZGS thanks D. Huse for several clarifying discussions.
SKS thanks DST-INSPIRE for financial support.

\section*{References}

%
\end{document}